\documentclass[fleqn,usenatbib]{mnras}

\usepackage{newtxtext,newtxmath}

\usepackage[T1]{fontenc}
\usepackage{stfloats}
\usepackage{graphicx}	
\usepackage{amsmath}

\title[ANT Dust Reprocessing Echoes]{Mid-Infrared Echoes of Ambiguous Nuclear Transients Reveal High Dust Covering Fractions: Evidence for Dusty Tori}

\author[Jason T. Hinkle]{
\href{http://orcid.org/0000-0001-9668-2920}{Jason T. Hinkle}$^{1}$\thanks{jhinkle6@hawaii.edu}
\\
$^{1}$Institute for Astronomy, University of Hawai`i, 2680 Woodlawn Drive, Honolulu, HI 96822, USA
}

\pubyear{2022}

\begin{document}
\label{firstpage}
\pagerange{\pageref{firstpage}--\pageref{lastpage}}
\maketitle

\begin{abstract}
Alongside the recent increase in discoveries of tidal disruption events (TDEs) have come an increasing number of ambiguous nuclear transients (ANTs). These ANTs are characterized by hot blackbody-like UV/optical spectral energy distributions (SEDs) and smooth photometric evolution, often with hard powerlaw-like X-ray emission. ANTs are likely exotic TDEs or smooth flares originating in active galactic nuclei (AGNs). While their emission in the UV/optical and X-ray has been relatively well-explored, their infrared (IR) emission has not been studied in detail. Here we use the NEOWISE mission and its low-cadence mapping of the entire sky to study mid-infrared dust reprocessing echoes of ANTs. We study 19 ANTs, finding significant MIR flares in 18 objects for which we can estimate an IR luminosity and temperature evolution. The dust reprocessing echoes show a wide range in IR luminosities ($\sim10^{42} - 10^{45}$ erg s$^{-1}$) with blackbody temperatures largely consistent with sublimation temperature of graphite grains. Excluding the two sources possibly associated with luminous supernovae (ASASSN-15lh and ASASSN-17jz), the dust covering fractions (f$_c$) for detected IR flares lie between 0.05 and 0.91, with a mean of f$_c$ = 0.29 for all ANTs (including limits) and f$_c$ = $0.38 \pm 0.04$ for detections. These covering fractions are much higher than optically-selected TDEs and similar to AGNs. We interpret the high covering fractions in ANT host galaxies as evidence for the presence of a dusty torus.
\end{abstract}


\section{Introduction}

Supermassive black holes (SMBHs) lie in the centers of essentially all massive galaxies \citep[e.g.,][]{magorrian98, kormendy13}. While measurements of orbiting stars and/or gas \citep{ford94, atkinson05, ghez05, gebhardt11} can reveal the presence of a SMBH and constrain important properties like mass, this is only possible for a small number of nearby galaxies. More commonly, the presence of a SMBH is inferred by the existence of a spatially unresolved, luminous nuclear source, otherwise known as an active galactic nucleus \citep[AGN; ][]{seyfert43, schmidt63, antonucci93}. In recent years, transient events such as tidal disruption events \citep[TDEs; e.g.,][]{holoien14a, holoien14b, gezari12b, gezari21} and AGN flares \citep[e.g.,][]{graham17, trakhtenbrot19a} have also provided key insight into the properties of SMBHs and their accretion behaviors.

Largely due to the expansion of transient surveys like the All-Sky Automated Survey for Supernovae \citep[ASAS-SN;][]{shappee14, kochanek17}, the Asteroid Terrestrial Impact Last Alert System \citep[ATLAS;][]{tonry18}, Gaia Alerts \citep{wyrzykowski12}, the Panoramic Survey Telescope and Rapid Response System \citep[Pan-STARRS;][]{chambers16}, and the Zwicky Transient Facility \citep[ZTF;][]{bellm19} the number of discovered nuclear transients is rapidly expanding. One notable surprise is an increasing number of transients with unusual properties, often showing features expected of distinct source classes. Such objects have been dubbed ambiguous nuclear transients (ANTs), with well-studied examples including PS16dtm \citep{blanchard17}, ASASSN-18jd \citep{neustadt20}, ASASSN-18el \citep{trakhtenbrot19b, ricci20, hinkle22a}, and ASASSN-20hx \citep{hinkle21c}. 

ANTs share characteristics of both TDEs and AGN flares, often showing a hot UV/optical blackbody-like SED with hard, powerlaw-like X-ray emission \citep[e.g.,][]{neustadt20, frederick21, hinkle21c}. Interestingly, many ANTs show relatively smooth photometric evolution \citep[e.g.,][]{trakhtenbrot19b, frederick21, hinkle21c}, straining the theoretical predictions of stochastic variability induced by instabilities in AGN accretion disks \citep[e.g.,][]{janiuk11}. While the physical mechanisms driving some ANTs have been claimed in the literature, typically a TDE or AGN flare, \citep[e.g.,][]{ricci20, frederick21}, many remain unclear \citep[e.g.,][]{neustadt20, hinkle22a}.

One hope for transients occurring on SMBHs is that properties of the black hole, such as mass \citep[e.g.,][]{mockler19, ryu20} and spin \citep[e.g.,][]{reynolds19, gafton19}, can be inferred from observations of these events. However, to understand the connection of transients to their SMBHs, we must first understand the environments of these SMBHs. In the case of quasi-static AGNs, significant work has been done to establish the so-called unified model \citep{antonucci93, urry95, netzer15}. In this model, the observed emission is set primarily by obscuration due to the dusty torus rather than differences in the intrinsic physical geometry. Nevertheless, the growing number of optical changing-look AGNs \citep[e.g.,][]{shappee14, denney14, lamassa15, neustadt23, guo24, zeltyn24} indicate that such a model cannot explain all AGN phenomena.

To begin to understand the physical environments in which transients occur, we must look across the electromagnetic spectrum. For example, observations of TDEs in the radio \citep[e.g.,][]{alexander20, cendes21} give constrains on the density profiles of nuclear gas. X-ray measurements of TDEs and large-amplitude AGN flares \citep[e.g.,][]{ricci20, yao22} can reveal the formation of coronae around SMBHs. In the infrared (IR), dust reprocessing echoes of TDEs give insight into the nuclear dust covering fractions \citep{vanvelzen16b, njiang16, lu16, jiang21b, cao22}, showing roughly 1\% covering fractions at sub-pc scales. Concurrently, the discovery of several obscured TDE candidates \citep[e.g.,][]{mattila18, kool20, onori22} shows that luminous nuclear flares clearly occur in dusty environments as well. Further constraints on the dust covering fraction for a growing sample of transients, enabled by the study of dust reprocessing echoes, can differentiate between transients happening the dust-poor and dust-rich environments. This may give important insights into the types of galaxies in which transients tend to occur.

Here we search for dust reprocessing echoes in a sample of ANTs using NEOWISE \citep{mainzer11} to constrain the properties of nuclear dust for such sources. In Section \ref{data} we detail our sample selection and data. In Section \ref{analysis} we present our analysis of the MIR light curves. We discuss our results in Section \ref{discussion} and  summarize our findings in Section \ref{summary}. Throughout the paper we assume a cosmology of $H_0$ = 69.6 km s$^{-1}$ Mpc$^{-1}$, $\Omega_{M} = 0.29$, and $\Omega_{\Lambda} = 0.71$ \citep{wright06, bennett14}.

\section{Sample and Data}\label{data}

\subsection{Sample Selection}

\begin{table*}
\centering
 \caption{Sample of ANTs}
 \label{tab:sample}
 \begin{tabular}{ccccccc}
  \hline
  Object & TNS ID & $z$ & log$\bigl(\frac{\textrm{M}_{\textrm{BH}}}{\textrm{M}_{\odot}}\bigr)$ & Right Ascension & Declination & References \\
  \hline
  ASASSN-15lh & SN2015L & 0.2326 & 8.3$^{b}$ & 22:02:15.451& $-$61:39:34.60 & \citet{dong16, leloudas16, godoy-rivera17} \\ 
  ASASSN-17cv & AT2017bgt & 0.064 & 7.3$^{c}$ & 16:11:05.696 & $+$02:34:00.52 & \citet{trakhtenbrot19a} \\
  ASASSN-17jz & AT2017fro & 0.164 & 7.5$^{d}$ & 17:19:55.850 & $+$41:40:49.48 & \citet{holoien21} \\
  ASASSN-18jd & AT2018bcb & 0.1192 & 7.6$^{e}$ & 22:43:42.871 & $-$16:59:08.49 & \citet{neustadt20} \\
  ASASSN-20hx & AT2020ohl & 0.0167 & 7.9$^{f}$ & 17:03:36.492	& $+$62:01:32.34 & \citet{hinkle21c} \\
  ASASSN-20qc & AT2020adgm & 0.056 & 7.3$^{a}$ & 04:13:02.45 & $-$53:04:21.72 & \citet{arcavi_20qc}, Pasham et al., in press \\
  ATLAS17jrp & AT2017gge & 0.066 & 6.6$^{g}$ & 16:20:35.004 & $+$24:07:26.57 & \citet{onori22, wang22} \\
  Gaia19axp & AT2019brs & 0.3736 & 7.2$^{h}$ & 14:27:46.400 & $+$29:30:38.27 & \citet{frederick21} \\
  Gaia21exd & AT2021acia & 0.296 & 8.6$^{a}$ & 00:51:39.960 & $-$30:24:25.52 & \citet{hodgkin_21acia, hinkle_21acia} \\
  iPTF16bco & \dots & 0.237 & 7.8$^{i}$ & 15:54:40.256 & $+$36:29:52.09 & \citet{gezari17a, frederick19} \\
  OGLE17aaj & \dots & 0.116 & 7.4$^{j}$ & 01:56:24.930 & $-$71:04:15.70 & \citet{gromadzki19} \\
  PS16dtm & AT2016ezh & 0.0804 & 6.0$^{k}$ & 01:58:04.739  & $-$00:52:21.74 & \citet{blanchard17} \\
  ZTF18aajupnt & AT2018dyk & 0.0367 & 5.5$^{i}$ & 15:33:08.015	& $+$44:32:08.20 & \citet{frederick19} \\
  ZTF18abjjkeo & AT2020hle & 0.103 & 6.4$^{h}$ & 11:07:42.871 & $+$74:38:02.16 & \citet{frederick21} \\
  ZTF19aaiqmgl & AT2019avd & 0.0296 & 6.1$^{h}$ & 08:23:36.767 & $+$04:23:02.46 & \citet{frederick21} \\
  ZTF19aatubsj & AT2019fdr &  0.2666 & 7.1$^{h}$ & 17:09:06.859 & $+$26:51:20.50 & \citet{frederick21} \\
  ZTF19abvgxrq & AT2019pev & 0.097 & 6.4$^{h}$ & 04:29:22.720	& $+$00:37:07.50 & \citet{frederick21} \\
  ZTF20aanxcpf & AT2021loi & 0.083 & 7.2$^{a}$ & 01:00:39.619 & $+$39:42:30.31 & \citet{graham_21loi, makrygianni23} \\
  ZTF20acvfraq & AT2020adpi & 0.26 & 7.5$^{a}$ &23:18:53.770 & $-$10:35:05.82 & \citet{chu_20adpi}, Hinkle et al., in preparation \\
  \hline
 \end{tabular}\\
\begin{flushleft} The 19 ANTs analyzed in this manuscript. The TNS ID is the identification given for objects reported on the Transient Name Server. References include the discovery papers and the appropriate transient discovery and/or classification reports. In general the SMBH mass estimates come from virial mass estimates for sources in known AGNs, or host-galaxy scaling relations. The SMBH masses newly computed in this work are indicated by the superscript $a$. The virial mass references are $c$: \citet{trakhtenbrot19a}, $h$: \citet{frederick21}, and $i$: \citet{frederick19}. The host-galaxy scaling mass references are $b$: \citet{wevers17}, $d$: \citet{holoien21}, $e$: \citet{neustadt20}, $f$: \citet{hinkle21c}, $g$: \citet{onori22}, and $j$: \citet{gromadzki19}.\end{flushleft}
\end{table*}

To construct our sample of ANTs, we consider smooth nuclear flares with either a tentative classification of the source as a TDE or AGN flare in the literature, or with a publicly available classification spectrum showing features similar to other ANTs. We restrict our sample to optically-selected events, excluding sources such as AT2017gbl \citep{kool20}, as infrared-selected transients have weak observed UV/optical emission and thus poor constraints on their UV/optical temperatures and luminosities, necessary to determine dust covering fractions. We include events discovered through 2021, although only the rise to peak in the mid-infrared will be seen for recently-discovered events. This yields 19 sources, detailed in Table \ref{tab:sample}. 

Some of these ANTs have claimed source classifications in the literature, including PS16dtm \citep[TDE in a Narrow Line Seyfert 1, ][]{blanchard17}, ATLAS17jrp \citep[TDE in a dusty environment, ][]{onori22, wang22}, and several from the sample of \citet{frederick21}. Even so, many of these sources could be explained by either TDEs or smooth AGN flares. Furthermore, as smooth AGN flares are rare and may be associated with TDEs in existing AGN hosts \citep[e.g.,][]{chan19}, we consider them all to be ANTs for the purpose of this study. This decision is supported by the analysis of \citet{auchettl18}, who showed that $<4$\% of coherently declining nuclear X-ray transients are likely to be powered by AGNs rather than TDEs.

Many of the ANTs in our sample have conflicting classifications, particularly those suggested to be associated with supernovae. Perhaps none is more noteworthy than ASASSN-15lh, which has been claimed by various groups as either the most luminous SLSN-I \citep[e.g.,][]{dong16} or the most luminous TDE \citep[e.g.,][]{leloudas16} to date. While the constraint for ASASSN-15lh on its separation from the nucleus is consistent with it being a bona fide nuclear transient, the possibility of it being a non-nuclear SLSN cannot be ruled out. Similarly, \citet{holoien21} find that ASASSN-17jz may be most consistent with a nuclear Type IIn supernova, while not ruling out the possibility of a smooth AGN flare. The rest of our sample has much clearer signs of being linked to transient accretion onto a SMBH, such as strong X-ray emission or broad emission lines \citep[e.g.,][]{trakhtenbrot19a, neustadt20, frederick21}. As such, for the remainder of this paper we shall denote ASASSN-15lh and ASASSN-17jz separately from the rest of our sample to indicate their potential association with luminous supernovae.

In this paper, we quantify the smoothness of the photometric evolution of an ANT through a cut on the fractional variability superimposed on top of the overall flare profile. We measure this by fitting a spline to the bolometric light curve of each ANT, implemented using the \textsc{scipy.interpolate.splrep} and associated \textsc{scipy.interpolate.splev} functions. To avoid overfitting and underestimating the true fractional variability, we used a cubic spline weighted by the inverse of the uncertainties employing the the maximum recommended smoothing parameter. These generally provide good descriptions of the ANT bolometric light curves. We then computed the mean absolute error of the measured light curve relative to the spline fit and divided by the peak UV/optical luminosity of the ANT to measure a fractional variability. Based on the UV/optical variability of luminous AGNs \citep[e.g.,][]{ulrich97, peterson01, padovani17}, we place our variability threshold at 15\%. For our sample of 19 ANTs, each has a fractional variability below 10\% and the full sample has a mean fractional variability of 4\% with a standard deviation of 3\%. Future theoretical explorations of ANT-like flares will provide better physical motivation for the selection of ANTs from the growing population of nuclear flares.

Additional observational and theoretical investigations are needed to fully understand the separation of ANTs from typical TDEs and AGN flaring behaviors. Nevertheless, the multi-wavelength similarities of the ANTs, including their IR behaviors studied in this work, are highly suggestive of a similar physical origin. Furthermore, the smooth flares and blackbody-like UV/optical SEDs are atypical of AGNs, indicating that the population of ANTs is not likely to be an extension of normal AGN variability.

\subsection{Physical Parameters of the ANT Host Galaxies}

The redshifts of the ANT host galaxies are taken from either the appropriate publication (shown in Tab. \ref{tab:sample}) or from the public classification spectrum on the Transient Name Server\footnote{\url{https://www.wis-tns.org/}} (TNS). As the SMBH on which these ANTs occur is the dominant driver of accretion rates and timescales, it is vital to estimate the SMBH masses accurately. In general, we adopted the mass estimate favored in the discovery paper, but have also considered dedicated efforts to measure the SMBH masses of transient hosts. If such measurements were unavailable in the literature we computed our own estimate from host-galaxy scaling relations \citep{mcconnell13, mendel14}. For the few sources with multiple mass estimates from different papers, we followed an order of preference of virial mass and then galaxy-SMBH scaling relations. It should be noted that for some objects, the dispersion between various mass measurements can be nearly an order of magnitude. As none of these mass estimates are direct measurements, they each have an uncertainty of $\sim 0.3-0.5$ dex \citep[e.g.,][]{mcconnell13, guo20} and we therefore treat them as rough estimates.

\subsection{NEOWISE Mid-Infrared Data}

Launched in December 2009, the Wide-field Infrared Survey Explorer \citep[WISE; ][]{wright10} mapped the entire sky in four mid-infrared (MIR) bands: W1 (3.4 $\mu m$), W2 (4.6 $\mu m$), W3 (12 $\mu m$) and W4 (22 $\mu m$). It was placed in hibernation mode in February 2011 upon the exhaustion of its cryogens, necessary to operate in the W3 and W4 bands. In October 2013, the WISE mission was renamed NEOWISE \citep{mainzer11} and revived with a primary goal of detecting potentially hazardous near Earth objects \citep{mainzer14}.

During the NEOWISE mission each patch of the sky is observed every six months, typically with a series of twelve 7.7 s exposures in the W1 and W2 bands over the course of roughly a day. We collected NEOWISE data for each of our ANTs from the single exposure catalog. We used the NEOWISE 2022 Data Release, and therefore MIR data is available for our sources through December 2021, with the next release expected in Spring 2023. As the timescale of dust reprocessing echoes is many months to years, we have stacked all the exposures within a given epoch to obtain deeper constraints. This yields a MIR light curve of each ANT between 2013 and 2021 at a cadence of 6 months, sufficient to search for dust reprocessing echoes \citep[e.g.,][]{jiang21a, vanvelzen21b}. Unfortunately, a small number of ANTs apparently have less data than is typical for an average point on the sky. This includes ASASSN-18el \citep{trakhtenbrot19b}, which has too few points to include in our sample, and ZTF19aatubsj \citep{frederick21}, which has a unusually short baseline prior to the IR outburst.

\subsection{Swift UV/optical data}

\begin{table*}
\centering
 \caption{UV/optical Blackbody Fits}
 \label{tab:UVopt_BBs}
 \begin{tabular}{cccccccccccc}
  \hline
  Object & TNS ID & MJD & log$\bigl(\frac{\textrm{L}}{\textrm{erg } \textrm{s}^{-1}}\bigr)$ & dlog(L$_{l}$) & dlog(L$_{u}$) & log$\bigl(\frac{\textrm{R}}{\textrm{cm}}\bigr)$ & dlog(R$_{l}$) & dlog(R$_{u}$) & log$\bigl(\frac{\textrm{T}}{\textrm{K}}\bigr)$ & dlog(T$_{l}$) & dlog(T$_{u}$) \\
  \hline
    ASASSN-15lh  & SN2015L &  57197.0  & 45.40  & 0.01  & 0.01 &  15.75  & 0.01 &  0.01 &  4.26 &  0.01 &  0.01 \\
    ASASSN-15lh &  SN2015L &  57199.8  & 45.37 &  0.01  & 0.01 &  15.72  & 0.02 &  0.01 &  4.27  & 0.01 &  0.01 \\
    ASASSN-15lh &  SN2015L &  57201.8 &  45.34 &  0.01  & 0.01 &  15.75 &  0.02 & 0.02 &  4.24  & 0.01 &  0.01 \\
    ASASSN-15lh &  SN2015L &  57205.5 &  45.30  & 0.01 &  0.01  & 15.76 &  0.02  & 0.02 &  4.23  & 0.01 &  0.01 \\
    ASASSN-15lh  & SN2015L  & 57208.6 &  45.23 &  0.01 &  0.01 & 15.74  & 0.02 &   0.02 &  4.23  & 0.01 &  0.01 \\
  \hline
 \end{tabular}\\
\begin{flushleft} Bolometric UV/optical luminosity, effective radius, and temperature estimated from blackbody fits to the host-subtracted and extinction-corrected Swift data. We only give data for sources not previously published in \citet{hinkle21b}. A small subset of the data for ASASSN-15lh is shown here to illustrate the format and the full table is available as an ancillary file. \end{flushleft}
\end{table*}

ANTs often show hot UV/optical blackbody SEDs \citep{neustadt20, hinkle21c}. In many cases, the MIR emission from this UV-dominated blackbody is small compared to the signal seen in the NEOWISE light curves. Nevertheless, we must properly account for the contribution of the transient blackbody itself to the W1 and W2 flux. This is important since for weak echoes the relative contribution can be large in some epochs. As is common for TDEs and ANTs \citep[e.g.,][]{holoien14b, hinkle21a, nicholl20, neustadt20, hinkle21c}, we used Swift UVOT \citep{roming05} data to estimate the effective temperature and luminosity of the transient in the UV/optical. 

Many of our sources have Swift UVOT data and blackbody fits available in \citet{hinkle21b}, which we adopt here. For several sources, we reduced available, unpublished Swift data and fit them as a blackbody following the methods of \citet{hinkle21b}. The results of these blackbody fits are given in Table \ref{tab:UVopt_BBs}. For each of the ANTs with blackbody fits to Swift data, we estimated a bolometric light curve by scaling their optical light curves to match the interpolated Swift bolometric luminosity, similar to previous TDEs/ANTs \citep{holoien20, holoien21, hinkle21c}. 

Three sources (iPTF16bco, ZTF18abjjkeo, and Gaia21exd) did not have multi-band Swift data near peak and therefore do not have reliable bolometric luminosity estimates. For these sources, we assumed a representative blackbody with a temperature of 20,000 K (roughly the median blackbody temperature of the ANTs in \citealt{hinkle21b}) and flat temperature evolution to scale the available optical light curves and estimate a bolometric light curve. For a conservative 25\% error on the blackbody temperature, this would correspond to a 0.2 dex uncertainty in the peak bolometric luminosity.

\section{Mid-Infrared Light Curve Analysis}\label{analysis}

\begin{figure*}
\centering
 \includegraphics[height=5.7cm]{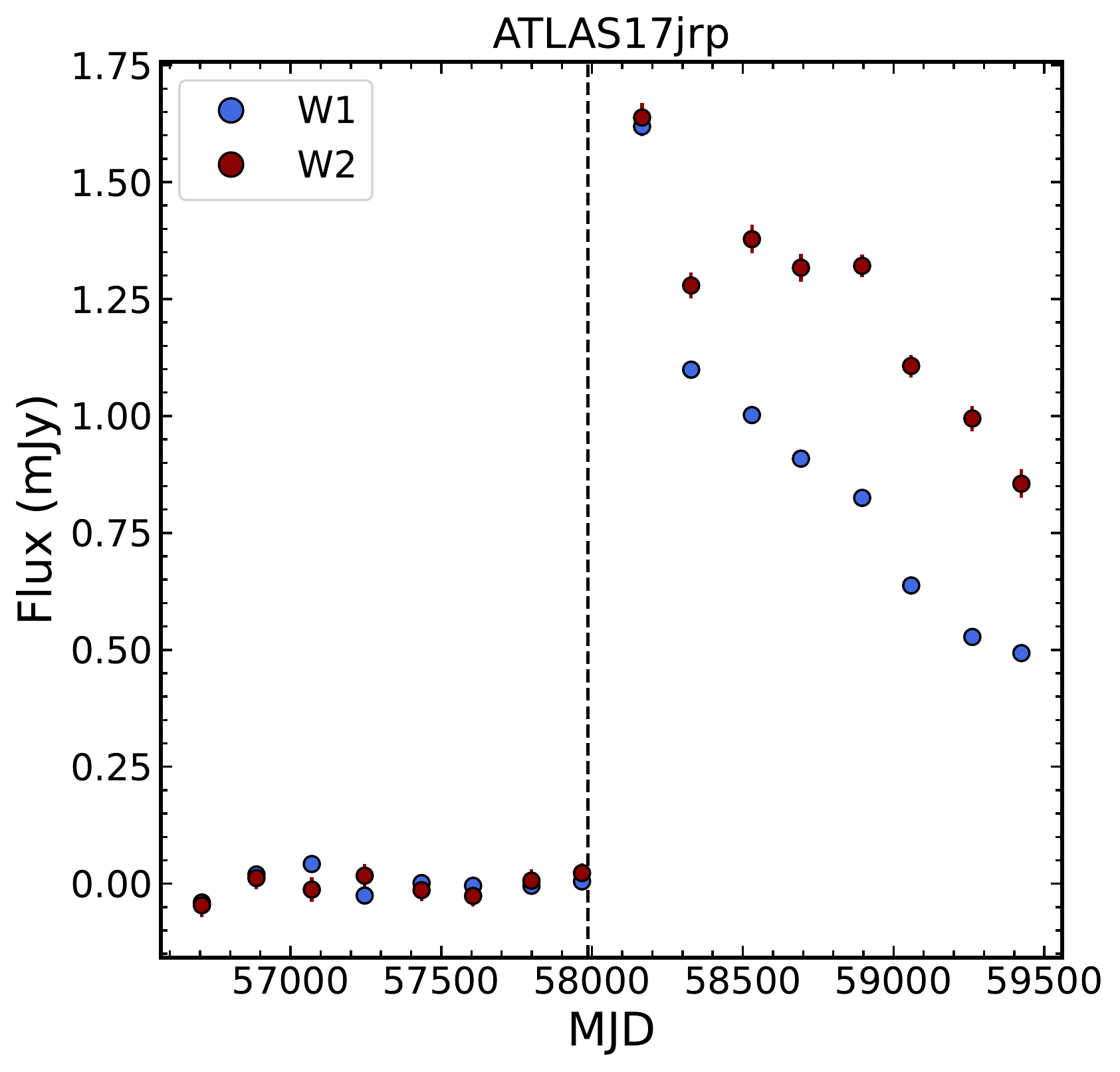}\hfill
 \includegraphics[height=5.7cm]{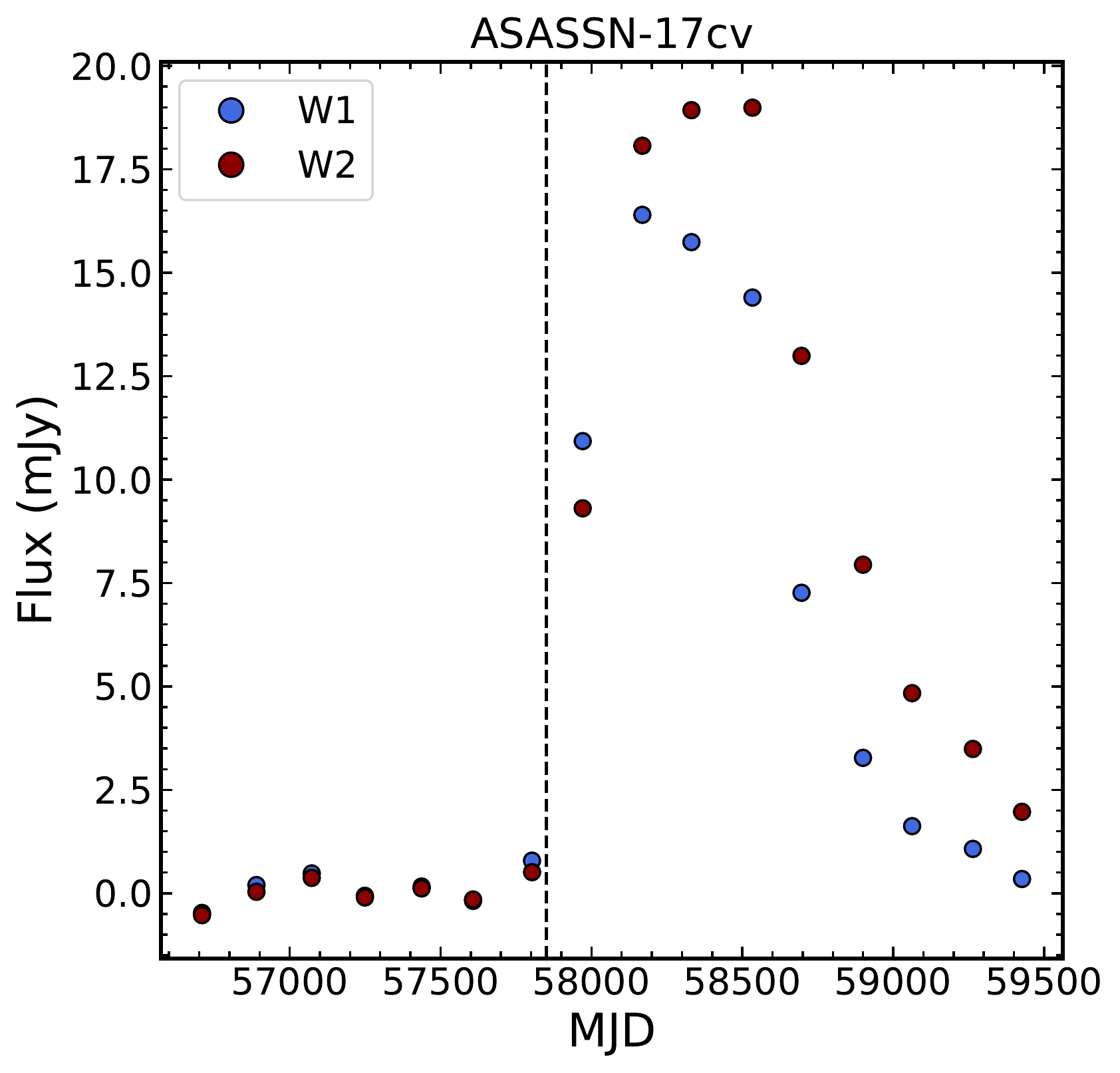}\hfill
 \includegraphics[height=5.7cm]{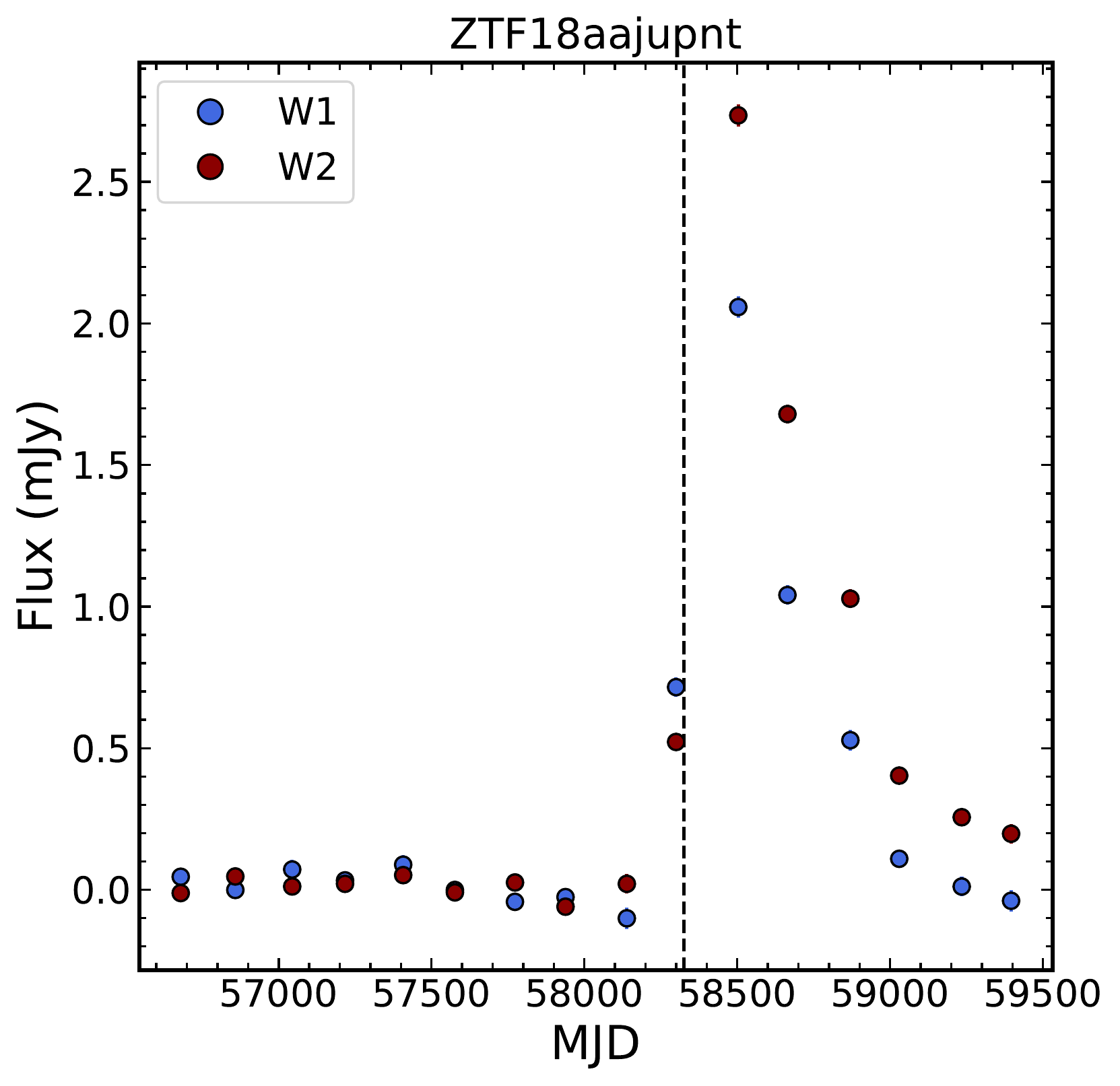}\hfill
 \caption{Example NEOWISE W1 and W2 light curves for the ANTs ATLAS17jrp (left panel), ASASSN-17cv (middle panel) and ZTF18aajupnt (right panel). The light curves have had the host and transient contribution subtracted and are corrected for Galactic foreground extinction. The vertical dashed lines represent the time of UV/optical peak. Each ANT dust reprocessing echo shows a rise beginning near the UV/optical peak and gets redder as the source evolves. Note that the flux scaling for each panel is different.}
 \label{fig:WISE_lcs}
\end{figure*}

\subsection{Stacking and Host Flux Subtraction}

The first step in our analysis of the NEOWISE MIR light curves was to stack the single exposures in a given epoch through a weighted average. As expected for such SMBHs, none of the sources show significant variability on these short timescales. 

Next, it was necessary to remove the host, or quiescent, contribution to the MIR flux. We estimated the host flux by fitting a flat line to the pre-outburst NEOWISE data in each band, equivalent to taking the weighted average. As the discovery dates for our ANTs cover several years, the total number of data points and amount of time included in this estimation of the host flux varies. We estimated the uncertainty in the host flux by summing the standard error on the flux in quadrature with the median single epoch flux uncertainty, to avoid unrealistically small ($<<$1\%) uncertainties on the host flux. This also balanced the fact that some sources had weakly variable pre-outburst light curves (likely stronger AGNs) whereas others were completely flat. This yielded median uncertainties of 0.02 and 0.03 mag in the W1 and W2 bands respectively. 

After estimating the host flux, we subtracted the light curves and corrected for Galactic foreground extinction \citep{schlafly11}. Several studies on dust reprocessing echoes have made use of image subtraction to obtain their light curves \citep[e.g.,][]{jiang21a}. Given the fact that the large majority of our ANTs show a strong MIR flare, minor changes in the host flux estimates will yield only small changes in the subtracted flux. The uncertainty on the host flux was added in quadrature with the photometric uncertainty in each epoch, although this only contributes significantly for sources where the flux of the flare is small compared to the host baseline flux. Table \ref{tab:host_mags} lists the estimated host magnitudes for each of our ANTs.

\begin{table}
\centering
 \caption{ANT Host Galaxy Magnitudes}
 \label{tab:host_mags}
 \begin{tabular}{cccccc}
  \hline
  Object & Points &  W1 Mag & W1 Err. & W2 Mag & W2 Err. \\
  \hline
  ASASSN-15lh & 3 & 17.44 & 0.03 & 17.77 & 0.06  \\ 
  ASASSN-17cv & 7 & 14.87 & 0.02 & 15.02 & 0.02\\
  ASASSN-17jz & 7 & 16.89 & 0.01 & 16.96 & 0.03\\
  ASASSN-18jd & 8 & 16.22 & 0.01 & 16.56 & 0.03 \\
  ASASSN-20hx & 17 & 13.78 & 0.01 & 14.53 & 0.01 \\
  ASASSN-20qc & 14 & 15.86 & 0.01 & 16.05 & 0.01\\
  ATLAS17jrp & 8 & 16.79 & 0.02 & 17.27 & 0.04 \\
  Gaia19axp & 11 & 17.05 & 0.02 & 16.81 & 0.03 \\
  Gaia21exd & 15 & 16.80 & 0.02 & 16.90 & 0.04 \\
  iPTF16bco & 3 & 16.67 & 0.03 & 16.67 & 0.05 \\
  OGLE17aaj & 6 & 17.69 & 0.02 & 18.02 & 0.06 \\
  PS16dtm &  5 & 17.93 & 0.04 & 18.23 & 0.13\\
  ZTF18aajupnt & 9 & 14.75 & 0.01 & 15.40 & 0.01 \\
  ZTF18abjjkeo & 12 & 16.79 & 0.01 & 17.09 & 0.03 \\
  ZTF19aaiqmgl & 10 & 15.46 & 0.01 & 16.04 & 0.02 \\
  ZTF19aatubsj & 1 & 16.51 & 0.01 & 16.58 & 0.03 \\
  ZTF19abvgxrq & 12 & 16.48 & 0.02 & 16.66 & 0.03 \\
  ZTF20aanxcpf & 15 & 16.54 & 0.02 & 16.51 & 0.03 \\
  ZTF20acvfraq & 14 & 18.60 & 0.08 & 18.39 & 0.27 \\
  \hline
 \end{tabular}\\
\begin{flushleft} Host galaxy magnitudes for the ANTs in our sample, presented in AB magnitudes. The number of points indicates the number of NEOWISE epochs fit to estimate the host magnitude. The number of points is lower for sources discovered close to the beginning of the NEOWISE coverage, discovered recently, or the few sources with data issues. \end{flushleft}
\end{table}

\subsection{Subtraction of the Transient Contribution}

Using the bolometric UV/optical light curves, we then estimated the contribution of the transient itself to the MIR emission at each epoch to isolate the emission from hot dust. At times when there was an estimate of the bolometric luminosity from the UV/optical data we simply linearly interpolated the luminosity and temperature of the transient at the NEOWISE epoch. Outside of the range of the UV/optical data, we constructed a model similar to the one used for some TDEs \citep{vanvelzen21, hammerstein22}. Prior to the first data point, we modeled the flare with a Gaussian rise. To better constrain the fits given the often low number of data points on the rise, we assumed a rise timescale of 100 days \citep[e.g.,][]{vanvelzen21}. After the last data point, we used an exponential decay model. We assumed a flat temperature evolution outside of the range of the bolometric light curve. We found that this model provided a reasonable estimate of the UV/optical luminosity during the NEOWISE coverage.

\begin{figure}
\centering
 \includegraphics[width=0.48\textwidth]{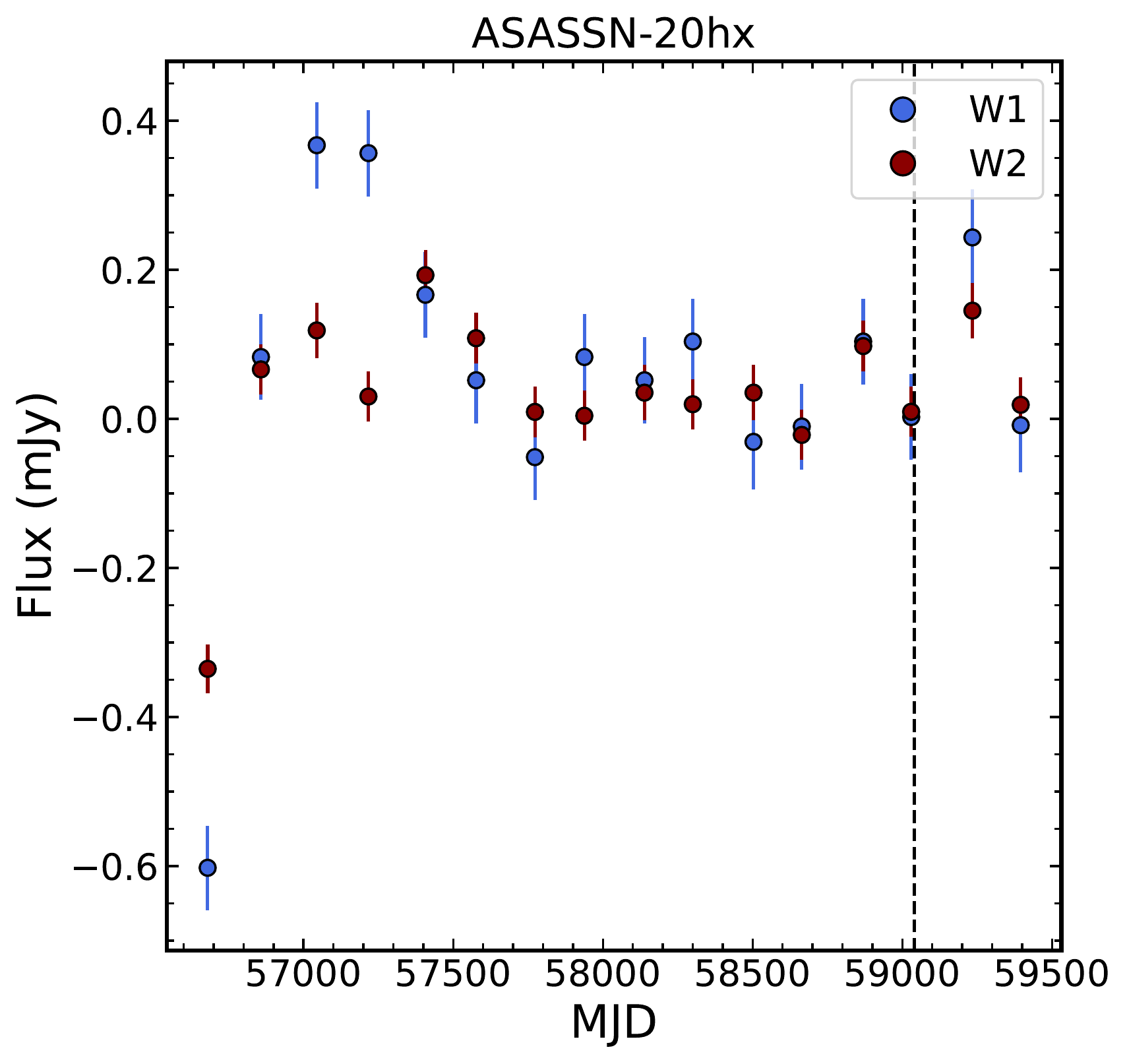}\hfill
 \caption{NEOWISE W1 and W2 light curves for ASASSN-20hx, the only ANT in our sample showing no clear dust reprocessing echo. The light curves have had the host and transient contribution subtracted and are corrected for Galactic foreground extinction. The vertical dashed line represents the time of UV/optical peak. The only detection after the optical flare began is similar in amplitude to the previous variability seen in the NEOWISE light curve.}
 \label{fig:20hx_lc}
\end{figure}

We then assumed a blackbody SED at each NEOWISE epoch with the interpolated blackbody temperature and scaled to the estimated bolometric luminosity.  We performed synthetic photometry at each epoch and scaled based on the host distance to estimate the W1 and W2 flux at the NEOWISE epochs, only subtracting the estimated transient contribution to the MIR flux after the beginning of the transient event itself. Figure \ref{fig:WISE_lcs} shows several example NEOWISE W1 and W2 band light curves with both the host and transient emission subtracted. These showcase both the typical behavior of these ANTs prior to peak, the diversity in rise times and slopes, as well as the general decrease in temperature as the dust reprocessing echo evolves.  In Figure \ref{fig:20hx_lc}, we highlight ASASSN-20hx, the only ANT to show no clear dust reprocessing echo. We note that there is weak variability prior to the flare and that the sole detection of an IR excess post-outburst is consistent with this underlying variability.

\subsection{Blackbody Fitting}

\begin{figure}
\centering
 \includegraphics[width=0.48\textwidth]{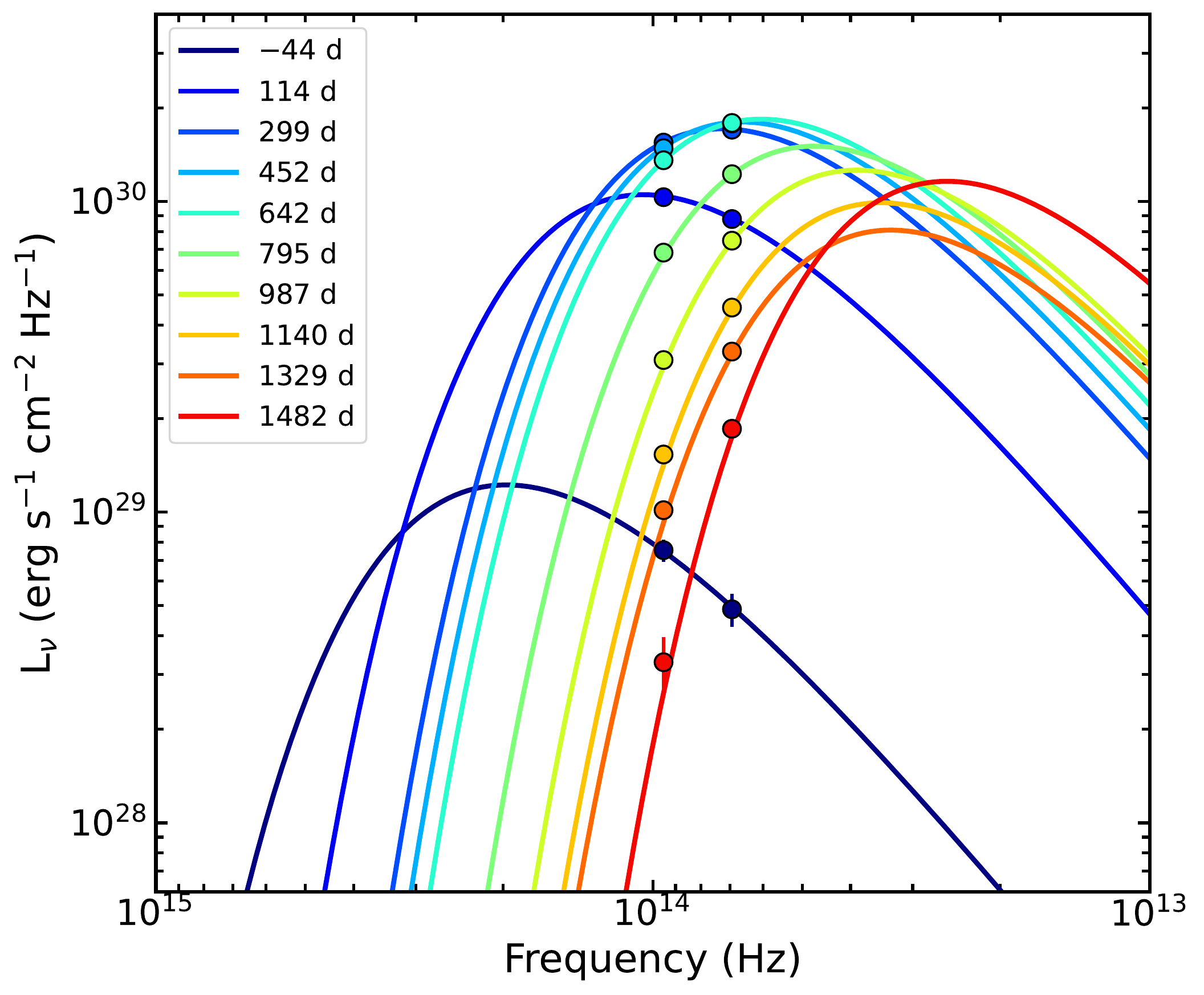}\hfill
 \caption{Evolution of the IR photometry (points) and corresponding median blackbody fits (lines) for the ANT dust reprocessing echo seen for ASASSN-17cv. The times shown in the legend are rest-frame days relative to the UV/optical peak.}
 \label{fig:17cv_BBs}
\end{figure}

\begin{figure*}
\centering
 \includegraphics[width=0.99\textwidth]{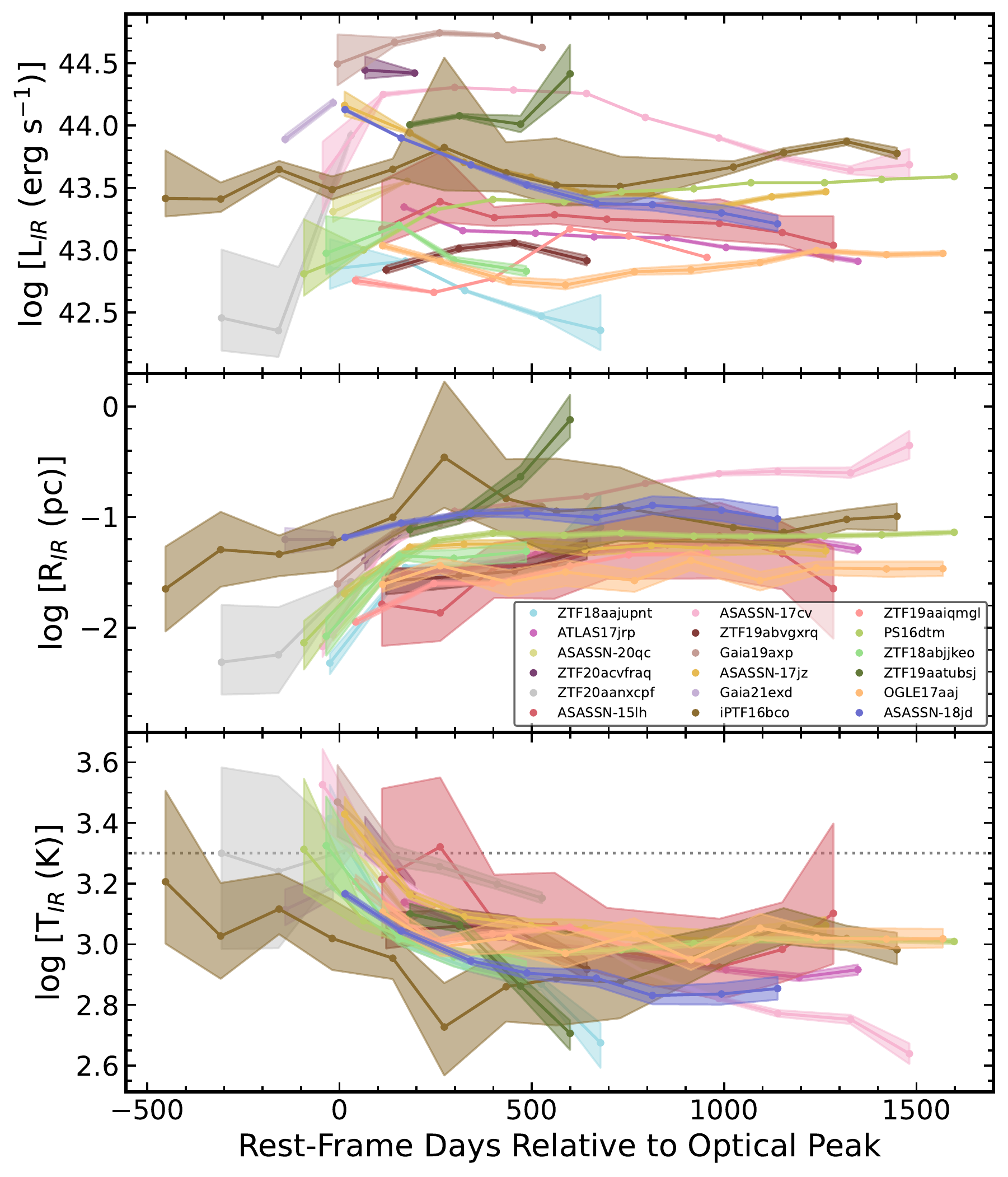}\hfill
 \caption{Evolution of the IR blackbody luminosity (top panel), effective radius (middle panel), and temperature (bottom panel) for the ANT dust reprocessing echoes studied in this work. The shading corresponds to the $1\sigma$ uncertainty. Time is in rest-frame days relative to the optical peak, except for the few sources discovered coming out of a seasonal break, for which time is relative to discovery. The dashed line in the bottom panel represents a conservative dust sublimation temperature limit of 2000 K, which the large majority of sources are below after the first epoch.}
 \label{fig:dust_BBs}
\end{figure*}

We fit each epoch of NEOWISE W1/W2 photometry as a blackbody using Markov Chain Monte Carlo (MCMC) methods and a forward modeling approach. We obtained the WISE W1 and W2 filter response functions from the Spanish Virtual Observatories Filter Profile Service \citep{rodrigo12}. Our fits were restricted to epochs with 2$\sigma$ detections in each band to ensure robust luminosity and temperature estimates. To keep our fits relatively unconstrained, we ran each of our blackbody fits with flat temperature priors of 100 K $\leq$ T $\leq$ 5000 K. An example of the IR evolution and median blackbody fits for the well-observed flare ASASSN-17cv is shown in Figure \ref{fig:17cv_BBs}. The WISE colors correspond directly to a blackbody temperature, which appears to fade monotonically starting shortly prior to peak UV/optical through nearly 1500 days post peak. Using the same methodology for each of our sources, Figure \ref{fig:dust_BBs} shows the dust luminosity, radius, and temperature evolution for our sample of ANTs, excluding the upper limit for ASASSN-20hx. For the sources iPTF16bco and ZTF20aanxcpf, we include several detections of dust emission prior to the optical flare, most likely consistent with low-level AGN variability.

The use of a simple blackbody fit to estimate the IR luminosity and dust temperature allows us to avoid making assumptions on the properties of the dust in these galaxies \citep[e.g.,][]{mathis77, draine84}. Dust often does not emit as a perfect blackbody, with varying emissivity as a function of wavelength \citep[e.g.,][]{draine84, laor93, kruegel03}. Estimates of the wavelength dependence of the absorption coefficient vary \citep[e.g.,][]{barvainis87, kruegel03, vanvelzen16b} and depend strongly on both the assumed particle size and composition \citep[e.g.,][]{jiang17}. Additionally, there is a large difference in the value of the absorption coefficient between different dust species in the few micron region \citep{laor93, kruegel03}, close to the expected peak emission for the hot ($\sim 1000$ K) dust expected in galactic nuclei \citep{barvainis87, mor09, mor12, jiang21a}. Furthermore, \citet{jiang21a} find that the IR luminosity between a perfect blackbody and modified blackbody are consistent (c.f. their Figure 4) and less than the dispersion in luminosities from different assumed dust properties. The dust temperature from a perfect blackbody often overestimates the temperature measured from a modified blackbody by $\sim 15\%$ \citep{jiang21a}. However, as our primary goal is to measure covering fractions, solely dependent on the peak IR luminosity, the simple blackbody approach is sufficient.

\begin{table}
\centering
 \caption{ANT Dust Covering Fractions}
 \label{tab:cover}
 \begin{tabular}{ccccc}
  \hline
  Object & f$_c$ &  f$_c$ Error & Type 1 f$_c$ & Type 2 f$_c$ \\
  \hline
  ASASSN-15lh & 0.008 & 0.003 & 0.21 & 0.25 \\
  ASASSN-17cv & 0.52 & 0.09 & 0.60 & 0.87 \\ 
  ASASSN-17jz & $>$0.20 & -- & 0.40 & 0.64 \\ 
  ASASSN-18jd & $>$0.31 & -- & 0.48 & 0.77 \\ 
  ASASSN-20hx & $<$0.05 & -- & 0.25 & 0.36 \\ 
  ASASSN-20qc & $>$0.25 & -- & 0.44 & 0.71 \\ 
  ATLAS17jrp & $>$0.17 & -- & 0.38 & 0.61 \\ 
  Gaia19axp & 0.50 & 0.08 & 0.59 & 0.86 \\ 
  Gaia21exd & $>$0.19 & -- & 0.39 & 0.63 \\ 
  iPTF16bco & 0.13 & 0.03 & 0.34 & 0.53 \\ 
  OGLE17aaj & $>$0.22 & -- & 0.41 & 0.67 \\ 
  PS16dtm & $>$0.13 & -- & 0.34 & 0.53 \\ 
  ZTF18aajupnt & 0.42 & 0.15 & 0.55 & 0.84 \\ 
  ZTF18abjjkeo & 0.15 & 0.03 & 0.35 & 0.57 \\ 
  ZTF19aaiqmgl & 0.91 & 0.31 & 0.74 & 0.96 \\ 
  ZTF19aatubsj & $>$0.46 & -- & 0.57 & 0.85 \\ 
  ZTF19abvgxrq & 0.045 & 0.005 & 0.25 & 0.35 \\ 
  ZTF20aanxcpf & $>$0.23 & -- & 0.42 & 0.68 \\ 
  ZTF20acvfraq & $>$0.28 & -- & 0.46 & 0.74 \\ 
  \hline
 \end{tabular}\\
\begin{flushleft} Dust covering fractions for the ANTs in the sample. The first column of covering fractions is computed through the ratio of IR and UV/optical peak luminosities as has been done for other transients. The columns labeled ``Type 1'' and ``Type 2'' use the corresponding AGN anisotropy corrections from Table 1 of \citet{stalevski16} for an aligned disk and torus with $\tau_{9.7} = 5$. Lower and upper limits are indicated in the first column as appropriate. \end{flushleft}
\end{table}

\subsection{Covering Fraction Measurements}

Following \citet{jiang21a}, we define the dust covering fraction (f$_c$) as the ratio of the peak IR luminosity to the peak UV/optical luminosity. To compute the covering fraction we compared the peaks of the bolometric UV/optical light curves and bolometric IR light curves. We estimated the peak of these light curves using \textsc{scipy.interpolate.splrep} and the associated \textsc{scipy.interpolate.splev}. In general, we fit the UV/optical data within 100 days of peak and all of the IR data with a high order spline to estimate the time and value of the peak luminosity in each band. While we typically used a 5th order spline for the UV/optical data and a 3rd order spline for the IR data, we confirmed that for well-behaved fits the results did not depend on the degree of the spline fit. For some ANTs, it was necessary to change these parameters. We checked each fit by eye to confirm that it well represented the peak in each wavelength regime.

Once we achieved a reasonable spline fit, we performed 1000 Monte Carlo iterations, perturbing the UV/optical and IR luminosities assuming Gaussian errors and refitting. We took the median value of this distribution to be the peak luminosity and added the one-sided $1\sigma$ uncertainties in quadrature to estimate the uncertainty on the peak luminosity. The covering fraction was then the ratio of the IR peak luminosity to the UV/optical peak luminosity with the uncertainties calculated through error propagation. The results are given in Table \ref{tab:cover}. Figure \ref{fig:covering_frac} shows the dust covering fractions for our sample of ANTs both as a function of redshift and SMBH mass. No notable trends are seen with either of these quantities. 

Amongst other parameters, \citet{frederick21} use the color (or temperature) evolution of a nuclear flare to help differentiate between more TDE-like and AGN-like behavior (c.f. their Figure 10), with TDEs often showing flat temperature evolution \citep[e.g.,][]{hinkle21b, vanvelzen21}. Therefore, in an attempt to separate ANTs into observational categories, we highlight ANTs with significant temperature evolution in Fig. \ref{fig:covering_frac} separately from sources showing no temperature evolution. To determine if a source had significant temperature evolution we used the Kendall Tau test \citep{knight66} for the UV/optical blackbody temperature as a function of time and selected sources with $\vert \tau \vert > 0.3$ as having significant temperature evolution. For the three sources with no UV/optical blackbody fits, we concluded there was no significant temperature evolution from the lack of evolution in their optical colors.

Many of our ANTs show lower limits on their covering fractions, with two main explanations. First, as in the case for PS16dtm and the sources discovered in 2020 or 2021, their IR flares are still rising and should continue to be visible in future NEOWISE releases. Second, some ANTs only show a declining MIR light curve, suggesting that we may have missed the peak. In several cases, the apparent lag between the UV/optical and IR peaks is short, indicating that the true covering fraction is likely not significantly higher than the lower limit. 

One ANT, ASASSN-20hx, shows no strong dust reprocessing echo, as can be seen in Fig. \ref{fig:20hx_lc}. There is a weak single detection after the discovery of the optical flare that is short-lived and comparable to previous variability seen in the MIR light curve. As such, it is unlikely that this is a dust reprocessing echo. Nevertheless, we treat the covering fraction implied by this epoch as an upper limit on the dust covering fraction.

\section{Discussion}\label{discussion}

An accurate picture of the environments of SMBHs is critical towards understanding how accretion flows originate and evolve, including the origins of the large-scale flaring behaviors studied here. These nuclear environments also play a large role in our ability to extract information about SMBHs from transient events. Nevertheless, such regions are nearly impossible to resolve, save for the closest galaxies \citep[e.g.,][]{ghez05, gravity20a, gravity21} or around luminous quasars \citep[e.g.,][]{gravity20b}. Transient events can help us understand the environments in which they occur by briefly illuminating them, as is the case for dust reprocessing echoes.

\subsection{Mid-Infrared Flare Characteristics}

\begin{figure*}
\centering
 \includegraphics[width=0.499\textwidth]{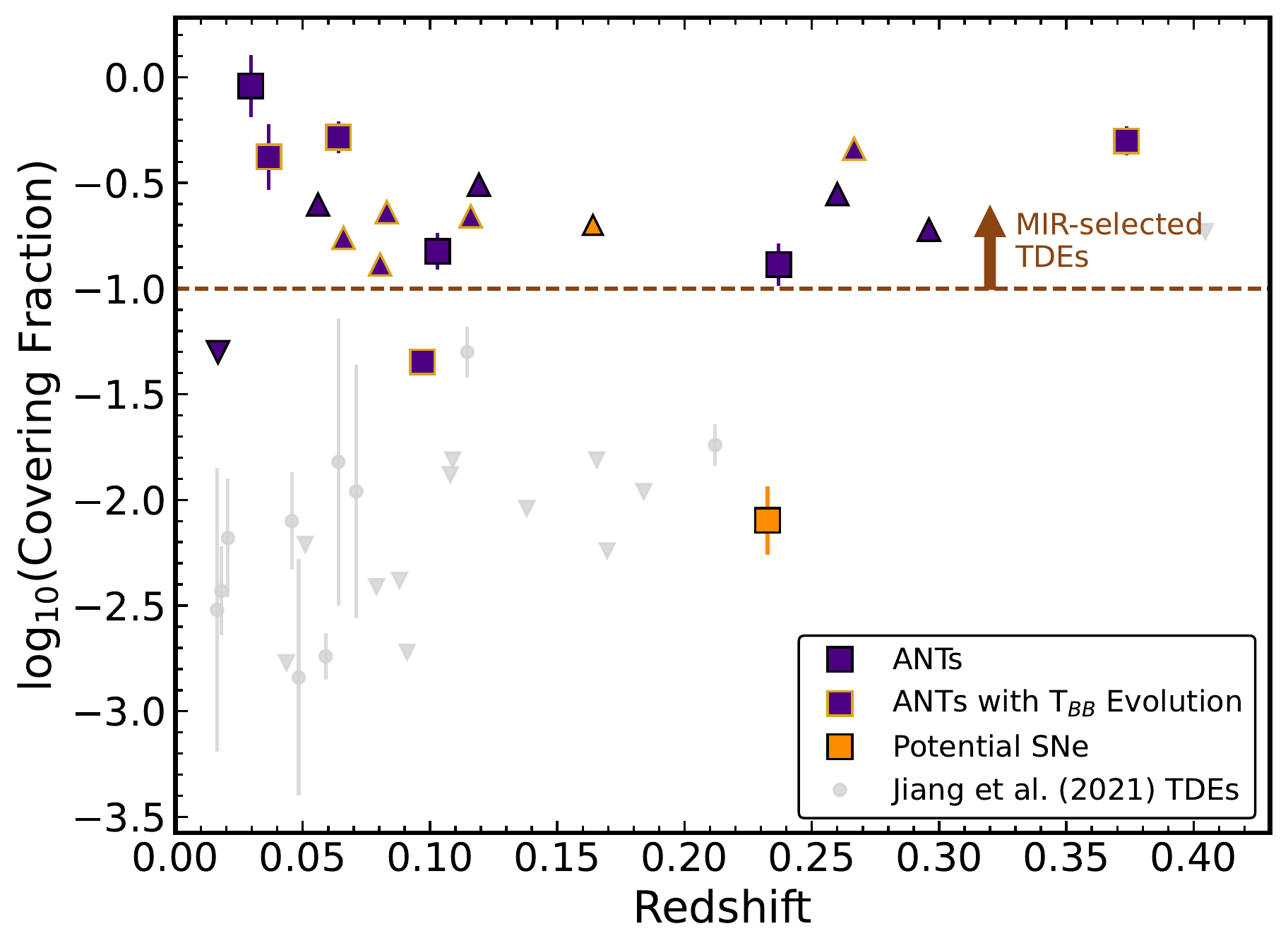}\hfill
 \includegraphics[width=0.499\textwidth]{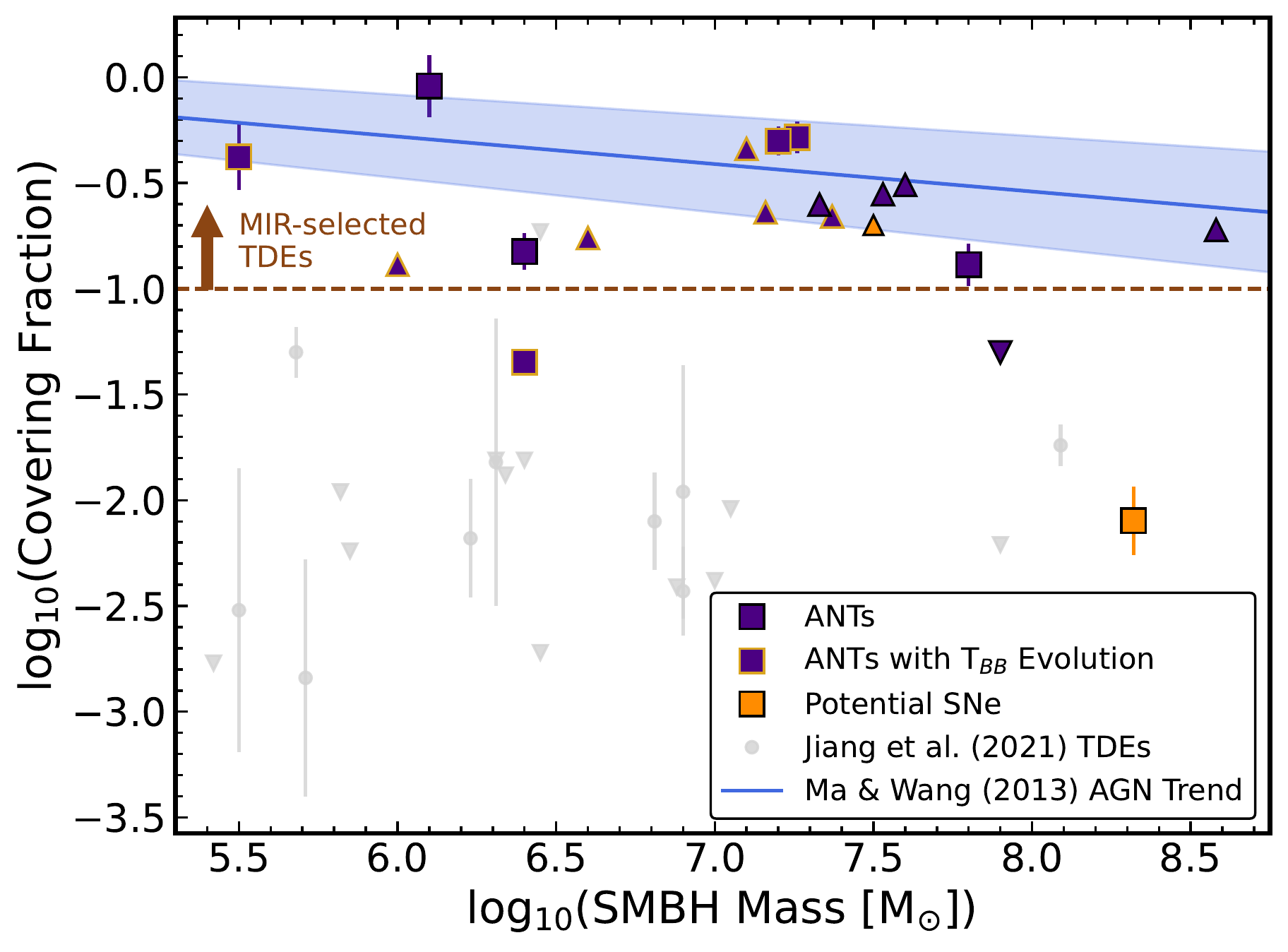}\hfill
 \caption{Dust covering fraction as compared to redshift (left panel) and SMBH mass (right panel). Squares mark detections, upward-facing triangles denote lower limits, and the downward-facing triangles represent upper limits. The ANTs are shown in purple, with a gold border indicating an ANT with significant temperature evolution. Potential SNe among the ANT sample are shown in orange. We exclude the uncertainties on SMBH mass as they are all of order $\sim 0.3-0.5$ dex and do not hold meaningful information about the relative confidence in the mass estimates. The translucent gray points are a comparison sample of TDEs from \citet{jiang21b}. The solid blue line is the best-fit trend between AGN dust covering fraction and SMBH mass from \citet{ma13}, with the shaded error representing the 90\% confidence interval on their linear fit. The dashed horizontal line marks the estimated minimum covering fraction for the MIR-selected TDEs from \citet{masterson24}.}
 \label{fig:covering_frac}
\end{figure*}

By examining NEOWISE light curves of 19 ANTs (with typical examples shown in Fig. \ref{fig:WISE_lcs}), we find that almost all show strong IR flares with timescales of several years, characteristic of reprocessing by nuclear dust \citep[e.g.,][]{vanvelzen16b, yang19, jiang21a, son22}. Most of our ANTs show little variability in NEOWISE prior to their IR flares, indicating at most weak AGN activity, comparable to analysis of archival data presented in several ANT discovery papers \citep{frederick19, hinkle21c}. The general shape and color evolution are similar to known MIR flares \citep{jiang21a}, but somewhat longer than the dust reprocessing echoes seen previously for TDEs \citep[e.g.,][]{vanvelzen16b, njiang16, jiang21b, vanvelzen21b}.

The ANT ASASSN-20hx \citep{hinkle21c} is the only source in our sample to not show a significant dust reprocessing echo. Shown in Fig. \ref{fig:20hx_lc}, the MIR light curve has only one high point after the optical flare and then immediately returns to the pre-flare level. Additionally, over the coverage of NEOWISE, the host of ASASSN-20hx shows low-level variability, consistent with a low-luminosity AGN \citep[as suggested in][]{hinkle21c}. The upper limit computed from the sole late-time detection of f$_c < 0.05$ is slightly higher than, but fully consistent with TDEs \citep{vanvelzen16b, jiang21b} and significantly lower than typical AGNs discussed later \citep[e.g.,][]{mor11, ma13}.

After isolating the dust contribution to the MIR emission, we were able to fit the dust emission as a blackbody. We find a large range of IR luminosities, spanning over 2 orders of magnitude (see the top panel of Fig. \ref{fig:dust_BBs}), fully consistent with recent discoveries of MIR flares \citep[e.g.,][]{jiang21a}. Assuming a roughly constant covering fraction, this would be expected given the similar range of input UV/optical luminosities for these ANTs \citep[e.g.,][]{hinkle21b}. Most ANTs in our sample show a smooth evolution in their IR luminosity, with only a small number (i.e. iPTF16bco, OGLE17aaj) exhibiting a re-brightening in the MIR.

In terms of effective radius, we see an early initial increase within the first $\sim 100$ rest-frame days after peak with a shallower increase at later times (see the middle panel of Fig. \ref{fig:dust_BBs}). This likely corresponds to the UV/optical light reaching the expected dust sublimation radius \citep[e.g.,][]{vanvelzen16b}. The sublimation radius for typical UV/optical flares calculated assuming graphite grains is $\sim 0.15$ pc, slightly larger than our late-time plateau. However, given a factor of $\sim 3$ underestimate of the dust radius for certain assumed dust properties as compared to a pure blackbody \citep{jiang21a}, our late-time radius flattening is consistent with the sublimation radius. Despite the large dispersion in the IR luminosity, the dispersion in effective radius for sources with very-late measurements ($t \sim 1000$ days) is small. 

The blackbody temperatures generally decrease over time, possibly plateauing at very late times relative to the UV/optical peak (see the bottom panel of Fig. \ref{fig:dust_BBs}). The initial temperature estimates for several sources are quite high, sometimes more than 2000 K, although often with large uncertainties. This is slightly higher than the sublimation temperature for graphite grains \citep[$\sim$$1800$ K; ][]{mor09, mor12, vanvelzen16b} and hotter than the sublimation temperature for silicate grains\citep[$\sim$$1500$ K; ][]{mor09, mor12}. This is likely due to the use of a pure blackbody model as compared to a modified blackbody, although the blue W1 $-$ W2 colors in many objects suggest that the dust may be quite hot at early times before the colors redden. Regardless, the hot temperatures suggest that the dust grains are more likely graphite in nature, as has been suggested for TDEs \citep[e.g.,][]{vanvelzen16b, vanvelzen21b}.  By a few hundred days after the UV/optical peak, the temperatures have settled to $\sim 700 - 1000$ K, well below the sublimation temperatures for either graphite or silicate grains. Similar to the effective radius, the temperatures have a small dispersion at late times.

From the ratio of the peak IR luminosity computed in this work to the peak UV/optical luminosity, either from the literature or determined here, we were able to estimate dust covering fractions for our ANT host nuclei. While the luminosity ratio may not always be an accurate proxy of the covering fraction due to the anisotropy of emission \citep{stalevski16}, the unknown emitting geometries of ANTs makes correcting for this effect difficult. Nevertheless, to illustrate these effects, we have applied the corrections found in Table 1 of \citet{stalevski16} for the case of an aligned disk and torus with $\tau_{9.7} = 5$ for both Type 1 and Type 2 AGN geometries. These updated covering fraction values are listed in Table \ref{tab:cover} along with our default covering fraction estimates with a methodology akin to other transient events. Further constraints on the geometry of the ANT emitting region are needed for more accurate estimation of the covering fraction, although the Type 1 AGN corrections of \citet{stalevski16} are likely more accurate for the ANTs in our sample than the Type 2 corrections given the observed UV/optical transients.

\subsection{The Landscape of IR Transients}

For the few sources that have had dust reprocessing echoes previously studied, such as ASASSN-15lh, PS16dtm, iPTF16bco, ZTF19aatubsj and ATLAS17jrp, we find excellent agreement between our results and published results. For ASASSN-15lh, \citet{jiang21b} find a median dust temperature of $\sim 1000$ K, a median IR luminosity of $\sim 3 \times 10^{43}$ erg s$^{-1}$, and a covering fraction of log(f$_c$) = $-1.83 \pm 0.22$. From our fits to ASASSN-15lh, we find a median temperature of  $\sim 1100$ K, a median IR luminosity of $\sim 2 \times 10^{43}$ erg s$^{-1}$, and a covering fraction of log(f$_c$) = $-2.10 \pm 0.16$. For PS16dtm, \citet{jiang17, jiang21b} report a covering fraction of log(f$_c$) > $-0.9$ as compared to our estimate of log(f$_c$) > $-0.89$. Additionally, the temperature evolution for the early-time PS16dtm epochs is in complete agreement with the results of \citet{jiang17}. For iPTF16bco, \citet{jiang21a} report an approximate covering fraction of f$_c$ = 0.1, fully consistent with our measurement of f$_c$ = $0.13 \pm 0.03$. For ZTF19aatubsj, our limit of f$_c$ > 0.46 is consistent with the previous estimate of f$_c \sim 0.33$ \citep{resuch22} given the additional epochs available in our study. Finally, for ATLAS17jrp, the estimate of f$_c \sim 0.2$ \citep{wang22} is consistent with our limit of f$_c > 0.17$.

The two potential SNe in our sample of ANTs, ASASSN-15lh and ASASSN-17jz, have distinct covering fractions. ASASSN-15lh shows only a very small f$_c$ = $0.008 \pm 0.003$ whereas ASASSN-17jz has a lower limit of f$_c > 0.20$. The dust luminosities for both objects are higher than the dust luminosities found for SLSNe \citep{sun22}. Additionally, no SLSNe have covering fractions higher than $\sim 0.1$ \citep{sun22}, and only a few percent in many cases. The higher covering fraction for ASASSN-17jz is difficult to explain solely with a SLSNe origin, except for the case in which the SN occurred in an AGN host nucleus, as suggested by \citet{holoien21}. Alternatively the covering fraction for ASASSN-17jz is consistent with it being an AGN flare.

As can be seen in Figure \ref{fig:covering_frac}, and excluding the two potential SNe, the majority of ANTs have covering fractions between 0.1 and 0.5. This is consistent with previous smaller samples of such transients \citep[e.g.,][]{vanvelzen21b}. Compared to the sample of TDEs presented in \citet{jiang21b}, the ANTs have significantly higher covering fractions, with the notable exception of ASASSN-15lh. A Kolmogorov–Smirnov test yields a probability of $7 \times 10^{-9}$ for all sources (including limits) and $8 \times 10^{-4}$ for detections only of the null hypothesis that the TDE and ANT covering fractions are drawn from the same distribution. Additionally, for typical corrections to the covering fraction calculated for AGN-like geometries \citep{stalevski16}, only one source (ZTF19aaiqmgl) would have its covering fraction decrease, with most ANTs still having higher covering fractions than TDEs. The ANT dust covering fractions and IR flares are strikingly similar to the IR-selected TDE sample of \citet{masterson24}. Such strong similarities imply that ANTs may be examples of TDEs occurring within an AGN host galaxy \citep[e.g.,][]{chan19, mckernan22, ryu24, wang24}.

In separating ANTs by their UV/optical blackbody temperature evolution \citep[e.g.,][]{frederick21}, we find no clear difference in their covering fractions. This may be explained by the fact that, regardless of the transient evolution itself, these ANTs likely occur in galaxies that host AGNs and therefore have AGN-like dust in their nucleus. We also investigated if there was any trend between the existence of strong (L$_{X} \gtrsim 10^{42}$ erg s$^{-1}$) X-ray emission and the dust covering fractions. Similarly, we find no clear trends, although there are relatively few ANTs without X-ray emission.

Finally, the measured properties of the ANT MIR flares are remarkably similar to the sample of MIRONGs \citep[Mid-infrared Outbursts in Nearby Galaxies,][]{jiang21a, wang22a, dodd23}. This includes similar distributions in the dust luminosity, radii, and temperatures. Much like the optical spectra of ANTs, the optical spectra of MIRONG hosts exhibit a range of emission line properties \citep{wang22a}. Notable similarities are Balmer emission line variability and the existence of coronal, \ion{He}{ii} $\lambda$4686, and/or Bowen emission features \citep{trakhtenbrot19a, neustadt20, onori22, wang22a}. The IR flares seen for ANTs are also similar to other IR-selected flare classes, including dust-obscured TDE candidates \citep[e.g.,][]{panagiotou23, masterson24}, and neutrino counterpart candidates \citep[e.g.,][]{resuch22, vanvelzen22}. Nevertheless, many IR-selected flares have such heavy dust obscuration that no optical transient emission is seen \citep{jiang21a, wang22}. Given these similarities, ANTs may provide important multi-wavelength insight into unseen UV/optical emission occurring in MIRONGs and other obscured accretion events.

\subsection{Comparison to AGN IR Emission}

Previous estimates of the covering fraction in AGNs vary from $\sim 0.1 - 0.7$ \citep[e.g.,][]{mor11, ezhikode17, ichikawa19, zhao21}. Similar covering fractions suggest that ANTs reside in galaxies that either host an AGN or otherwise have significant nuclear dust, unlike the post-merger systems preferred by TDEs \citep[e.g.,][]{french16, law-smith17}. Nevertheless, the relatively modest SMBH masses on which many ANTs occur may suggest that some ANTs are TDE-induced transients in dusty environments \citep{vanvelzen22}. However the ANTs with low covering fractions, like ASASSN-15lh, clearly reside in less dusty environments. This may indicate that they occurred outside of their host nucleus but we were unable to resolve the separation.

Similar to studies on hot dust in AGNs, we see no correlation between the covering fraction and redshift \citep[e.g.,][]{lusso13}. A correlation between covering fraction and SMBH mass has been claimed in some studies \citep[e.g.,][]{ma13} but not in others \cite[e.g.,][]{mor11}. In our sample of ANTs, we see no evidence for a correlation between the covering fraction and SMBH mass. Nonetheless, many ANTs lie close to the locus of covering fractions expected when extrapolating the linear trend of \cite{ma13} to lower SMBH masses (see the right panel of Fig. \ref{fig:covering_frac}). Several studies of strong AGNs have suggested an anti-correlation between covering fraction and either AGN bolometric luminosity \citep[e.g.,][]{maiolino07, mor11, alonsoherrero11, ma13} or Eddington ratio \citep[e.g.,][]{ezhikode17}. For low-luminosity AGNs however, studies suggest that the torus may disappear below an AGN bolometric luminosity of $\approx 10^{42}$ erg s$^{-1}$ \citep[e.g.,][]{elitzur06, hoenig07}, with the mechanisms powering their IR emission remaining unclear \citep{dumont20}. Many ANTs seem to occur in low-luminosity AGN hosts \citep[e.g.,][]{gezari17a, neustadt20, hinkle21c}, potentially in contrast with the expected covering fractions for AGNs with the lowest bolometric luminosities. However, the observed ANT covering fractions are generally consistent with the expectations for AGNs as a function of SMBH mass \citep[e.g.,][]{ma13} and several ANTs show high ($\gtrsim 0.5$) covering fractions, suggesting no conflict with previous studies on hot dust in AGNs.

Despite the high covering fractions, the dust extinction along our line of sight is likely relatively low. Almost all ANTs show blue UV/optical colors \citep[e.g.,][]{neustadt20, frederick21, hinkle21b} indicating low intrinsic extinction. Additionally, while 10 of 19 ANTs in our sample show some evidence for \ion{Na}{i} D absorption lines in their spectra near peak, indicating some dust along the line of sight \citep{poznanski12}, several sources with high covering fractions (e.g., Gaia19axp and ASASSN-17cv) do not have obvious \ion{Na}{i} D absorption features. Surprisingly ASASSN-20hx, with only an upper limit on the covering fraction, shows clear \ion{Na}{i} D absorption. Both cases are consistent with a clumpy nuclear dust distribution \citep[e.g.,][]{nenkova08}.

Given the similarity of the covering fractions to typical AGNs, we interpret the high dust covering fractions for our ANTs as evidence of dusty tori in the nuclei of ANT host galaxies. This further supports a link between AGN activity, albeit potentially weak AGN activity, and ANTs. The need for energy injection to support an extended torus \citep[e.g.,][]{pier92, thompson05, krolik07} and provide large covering fractions suggests that AGNs in ANT hosts are either still active or only turned off shortly befored the ANT occurred. This is not surprising as most ANTs show some signs of previous activity typically seen in narrow emission lines \citep[e.g.,][]{blanchard17, frederick21, holoien21} or X-ray properties \citep{trakhtenbrot19a, hinkle21c}.  Still, ANTs lie outside the typical behavior of AGNs and their exact physical mechanisms, though likely varied, remain unknown. Nevertheless, adding important constraints on their host environments, such as the evidence for dusty tori presented here will continue to shed light on this extreme form of SMBH activity.

\section{Summary}\label{summary}

In this manuscript we have collected NEOWISE light curves for a sample of 19 ANTs and isolated the emission from hot dust processing and re-emitting the UV/optical light emitted by the ANT itself. We fit this emission as a blackbody and computed dust covering fractions. Our main results are as follows.

\begin{itemize}

\item All but one of our ANTs (ASASSN-20hx) shows a strong MIR flare in NEOWISE consistent with a dust reprocessing echo. Several of the recently-discovered flares are still rising in the IR and will be interesting targets for continued follow-up.

\item The dust IR luminosities exhibit a wide range, between $10^{42.4}$ erg s$^{-1}$ and $10^{44.7}$ erg s$^{-1}$. The roughly two order of magnitude range in IR luminosities is similar to the observed range of UV/optical luminosities for ANTs.

\item The dust blackbody radii increase rapidly early on, reaching of order the expected sublimation radius for graphite grains by roughly 100 rest-frame days after the UV/optical peak and plateau thereafter.

\item The dust blackbody temperatures decrease over time, with a rapid decrease over the first several hundred rest-frame days after the UV/optical peak. While the initial dust temperatures may be above the sublimation temperature for graphite grains for some ANTs, all dust reprocessing echoes reach temperatures below $\sim 1000$ K at late times.

\item The dust covering fractions, computed as the ratio of the peak IR to peak UV/optical luminosity, range from 0.05 to 0.91 for detections (excluding the hotly contested source ASASSN-15lh), with a mean of $\sim 0.3 - 0.4$ depending on whether or not limits are taken into account. This is an order of magnitude higher than typical TDE dust covering fractions and similar to hot dust in AGNs. We interpret the high dust covering fractions as evidence for dusty tori in these ANT hosts.

\end{itemize}

Through the exploration of reprocessing echoes from hot dust in the nuclei of ANT host galaxies, it is clear that most ANTs occur in moderately dusty environments. While the paucity of dust in TDE hosts has been suggested to be evidence for an observational bias, these ANTs are selected by similar means to optically-selected TDEs. Regardless, it seems likely that ANTs occur in galaxies that either currently host an AGN, recently hosted an AGN, or otherwise have significant dust in the nuclei, possibly associated with large epochs of star formation. 

The NEOWISE mission will continue until at least June 2023, with its successor, the Near-Earth Object Surveyor, slated for launch in early 2026. This will allow us to study transients and their environments for many years to come. In particular, as the number of ANTs increases, we will continue to learn more about the range of extreme accretion events on SMBHs. While the UV/optical and X-ray will continue to give detailed insight into the properties of the transient events, MIR studies of dust reprocessing echoes will allow us to expand our understanding of the environments in which these elusive transients occur. 

\section*{Data availability}
	
The data underlying this article are available at the NASA/IPAC Infrared Science Archive and the NASA High Energy Astrophysics Science Archive Research Center .

\section*{Acknowledgements}

We thank the anonymous referees for comments that helped us improve the manuscript. We also thank Benjamin Shappee, Sjoert van Velzen, and Marko Stalevski for helpful discussions. We also thank Benjamin Shappee, Alexa Anderson, Dhvanil Desai, Christopher Storfer, and Samuel Walker for useful comments on the manuscript. JTH is supported by NASA award 80NSSC22K0127.

\bibliographystyle{mnras}
\bibliography{bibliography} 

\bsp	
\label{lastpage}
\end{document}